\documentclass[traditabstract]{aa} 
\usepackage{graphicx}
\usepackage{natbib}
\usepackage{longtable}
\newcommand{\HI}{H\,{\sc i}}
\newcommand{\skms}{\ensuremath{\,\mbox{km}\,\mbox{s}^{-1}}}

\usepackage{txfonts}
\begin{document}
\title{Lopsidedness in WHISP galaxies}

\subtitle{II. Morphological lopsidedness}

\titlerunning{Lopsidedness in WHISP galaxies: II. Morphological lopsidedness}

\author{J. van Eymeren
          \inst{1}
          \and
          E. J\"utte\inst{2}
          \and
          C.~J. Jog\inst{3}
          \and
           Y. Stein\inst{2}
	  \and
           R.-J. Dettmar\inst{2}
          }

 \institute{Fakult\"at f\"ur Physik, Universit\"at Duisburg-Essen, Lotharstr. 1, 47048 Duisburg, Germany\\ 
\email{janine.vaneymeren@uni-due.de}
         \and
            Astronomisches Institut der Ruhr-Universit\"at Bochum, Universit\"atsstr. 150, 44780 Bochum, Germany \\
             \email{eva.juette@astro.rub.de}
	 \and
	    Department of Physics, Indian Institute of Science, Bangalore 560012, India \\
             \email{cjjog@physics.iisc.ernet.in}
             }

   \date{Accepted 19 March 2011}

\abstract{
The distribution of stars and gas in many galaxies is asymmetric. This so-called lopsidedness is expected to significantly affect the dynamics and evolution of the disc, including the star formation activity. Here, we measure the degree of lopsidedness for the gas distribution in a selected sample of 70 galaxies from the Westerbork \HI\ Survey of Spiral and Irregular Galaxies. This complements our earlier work (Paper I) where the kinematic lopsidedness was derived for the same galaxies. The morphological lopsidedness is measured by performing a harmonic decomposition of the surface density maps. The amplitude of lopsidedness $A_1$, the fractional value of the first Fourier component, is typically quite high (about 0.1) within the optical disc and has a constant phase. Thus, lopsidedness is a common feature in galaxies and indicates a global mode. We measure $A_1$ out to typically one to four optical radii, sometimes even further. This is, on average, four times larger than the distance to which lopsidedness was measured in the past using near-IR as a tracer for the old stellar component, and will therefore provide a new, more stringent constraint on the mechanism for the origin of lopsidedness. Interestingly, the value of $A_1$ saturates beyond the optical radius. Furthermore, the plot of $A _1$ \emph{vs.} radius shows fluctuations which we argue are due to local spiral features. We also try to explain the physical origin of this observed disc lopsidedness. No clear trend is found when the degree of lopsidedness is compared to a measure of the isolation or interaction probability of the sample galaxies. However, this does not rule out a tidal origin if the lopsidedness is long-lived. Additionally, we find that the early-type galaxies tend to be more morphologically lopsided than late-type galaxies. Both results together indicate a tidal origin for the lopsidedness.
}

\keywords{Surveys -- Galaxies: evolution -- Galaxies: ISM -- Galaxies: structure}

\maketitle
%

\section{Introduction}
It is now known that both the stellar and the gas distribution in galaxies are often "lopsided". This feature was first highlighted in the classic paper by \citet{Baldwin1980}. \citet{Rix1995} analysed near-IR images of a sample of galaxies and characterised lopsidedness by the $m=1$ mode of a Fourier analysis. Since then, studies of larger samples of galaxies have been carried out that show that the fraction of lopsided galaxies is quite high and can be 50\,\%\ or more \citep[][]{Richter1994,Matthews1998,Bournaud2005,Angiras2006,Angiras2007}. Thus, lopsidedness is common, yet its properties are not as well-studied as the other common asymmetries such as spiral arms or bars which are characterised by the $m=2$ mode.

Lopsidedness can have a significant influence on the evolution of the host galaxy \citep[for a review of lopsided galaxies see e.g.,][]{Jog2009}.
In particular, since the $m=1$ mode has no Inner Lindblad Resonance, this mode allows the transport of matter to the central region \citep{1994A&A...288..365B}. Thus, lopsidedness can affect the star formation, and plays a potential role for fuelling the nuclear region (e.g., an AGN) \citep{2008ApJ...677..186R}. 

The high fraction of galaxies showing lopsidedness indicates that it is sustainable over a large period of time. However, its physical origin is still poorly understood. Several mechanisms might be responsible for the observed lopsided discs. Possible scenarios include tidal interactions and minor mergers \citep{Jog1997,1997MNRAS.292..349S,Zaritsky1997,2005A&A...437...69B}, asymmetric gas accretion \citep{Bournaud2005,Mapelli2008}, ram pressure from the intergalactic medium \citep{Mapelli2008}, but also an offset of the stellar disc in a halo potential \citep{Noordermeer2001}.

Recent studies revealed that lopsided is correlated with some properties of the galaxies. For instance, studies of \citet{Angiras2006} and \citet{Angiras2007} indicate that the fraction of lopsidedness depends on the environment. A high density group environment seems to favour the occurrence of lopsided discs. Furthermore, \citet{Bournaud2005} found a correlation with galaxy type, with late-type discs being more frequently lopsided than early-types. In contrast, the group studies of \citet{Angiras2006} show that the early-type galaxies are more lopsided than late-types which indicates tidal encounters as a likely mechanism for the origin of lopsidedness.

The physical origin of lopsidedness and its life time are interrelated and not yet well understood. A simple kinematic model gives a lifetime of $< 10^9$ years \citep{Baldwin1980}. A tidal origin which accounts for the self-gravity of the disc gives a somewhat longer time of 1-2 Gyr \citep{Bournaud2005}. A global mode treatment implies lopsidedness to be very long-lived, about the Hubble time \citep{Saha2007}, although a non-zero pattern speed would make it less long-lived \citep{Jog2009}.

In order to put some constraints on the physical origin of lopsidedness, we here present a new study of a large sample of disc galaxies. Thanks to several large \HI\ surveys carried out in the last years, the number of available and useful \HI\ maps has been increased enormously. We use deep medium-resolution \HI\ maps of 70 galaxies from the Westerbork \HI\ Survey of Irregular and Spiral Galaxies (WHISP\footnote{http://www.astro.rug.nl/$\sim$whisp/}). The advantage of using \HI\ instead of near-IR observations is that lopsidedness can be measured out to much larger radii, where lopsidedness is more pronounced \citep{Rix1995}. In fact, due to the low column density limit of the WHISP data, we have been able to measure the lopsidedness beyond the optical radius in almost all 70 galaxies. 

In a previous paper \citep[][hereafter Paper\,I]{vanEymeren2011}, the rotation curves of all sample galaxies are presented and the kinematic lopsidedness is analysed by comparing receding and approaching sides. Here, we complement the results from this kinematic analysis by a study of morphological lopsidedness at all radii.

This paper is organised as follows: in Sect.~2, we briefly mention the sample selection (a more detailed description can be found in Paper\,I). In Sect.~3, the harmonic analysis is explained. The results are presented in Sect.~4, which is followed by a discussion in Sect.~5. We summarise the main results and conclusions in Sect.~6.
\section{Sample selection}
For the purpose of this paper we selected 70 disc galaxies from the WHISP survey. We applied two selection criteria: (1) the galaxies need to have inclinations between 20\degr\ and 75\degr, otherwise, the tilted-ring analysis will not work properly; (2) the ratio of the \HI\ diameter over the beam size has to be larger than 10 in order to measure lopsidedness out to large radii. In order to combine reasonably high spatial resolution with sufficiently high signal to noise, we worked on data cubes smoothed to a resolution of 30\arcsec\,$\times$\,30\arcsec. As Fig.~1 in Paper\,I shows, the selection criteria do not bias our sample: the galaxies are distributed over a large range of \emph{B} magnitudes and morphological types. They typically extend out to one to four optical radii, sometimes even further.

A more detailed description of the selection process and some general properties of the final sample galaxies are given in Paper\,I. Details about the WHISP survey and the data reduction process can be found in \citet{Swaters2002}.
\section{Harmonic analysis}
The data analysis is based on routines within the Groningen Image Processing System \citep[GIPSY\footnote{URL: http://www.astro.rug.nl/$\sim$gipsy/},][]{vanderHulst1992}. A tilted-ring analysis to derive the kinematic parameters has already been performed in Paper\,I. We here concentrate on describing the harmonic decomposition, which gives us quantitative values of the morphological lopsidedness, and the calculation of the halo perturbation.
\subsection{Morphological lopsidedness of the \HI\ gas}
\label{secmorpholop}
The amplitude of lopsidedness in the morphology was calculated by decomposing the \HI\ surface density map into Fourier components:
\begin{equation}
\sigma(R,\phi)=a_0(R)+\sum a_m(R)cos(m\phi-\phi_m(R))
\label{eqsigma}
\end{equation}
with $a_0(R)$ being the mean surface density at a given radius \emph{R}, $a_m$
being the amplitude of the surface density harmonic coefficient, $\phi$ being
the azimuthal angle in the plane of the galaxy, and $\phi_m$ being the phase
of the $m^{th}$ Fourier coefficient. This harmonic decomposition was carried
out with the GIPSY task \emph{reswri} by using the best-fitting kinematic
parameters as derived in Paper\,I. Note that all pixels along a ring are
treated equally, which means that any small-scale structure is averaged out. The amplitudes were obtained up to the third Fourier component. We then calculated the normalised harmonic coefficients $A_m=a_m/a_0$. $A_1$ represents the normalised amplitude of the $m=1$ Fourier component denoting lopsidedness, whereas $A_2$ represents the amplitude of an $m=2$ component denoting a bar or a two-armed spiral feature.

It is crucial to accurately determine the dynamic centre and keep it (and the systemic velocity) fixed for all rings because if $x_0$, $y_0$, and $v_{\rm sys}$ are allowed to vary, the $m=2$ harmonic coefficients tend to rearrange themselves in order to minimise the effects of lopsidedness. In Paper\,I, we showed that the optical and dynamic centre usually agree within one beam size.
\subsection{Lopsidedness due to a perturbed halo}
\label{SectPerturbedpot}
We assume that the asymmetry arises due to the response of the disc to a halo distorted by a tidal encounter \citep{Jog1997}. Then, the perturbation potential can be obtained from the \HI\ surface density maps using the normalised amplitudes $A_1$, $A_2$, and $A_3$ for $m=1$, $m=2$, and $m=3$ respectively as obtained for an exponential disc \citep{Jog2000}.

In agreement with \citet{Angiras2006} we found that the \HI\ radial surface density profiles were in most cases roughly Gaussian in cross section. Therefore, we derived  the Gaussian scale length for all sample galaxies: first, a radial surface density profile was obtained from the \HI\ surface density maps. We then fitted a Gaussian curve of $S_0\,\exp(-(R-b)^2/2R^2_{\rm w})$ to the profiles using a $\chi^2$ fitting technique (GIPSY task \emph{gauprof}).

The relation between $\epsilon_1$, the perturbation parameter for the lopsided potential and $A_1$ for a radial Gaussian distribution \citep[see][]{Angiras2006} is given by 
\begin{equation}
\epsilon_1=\frac{A_1(R)}{2\left(\frac{R}{R_{\rm w}}\right)^2-1},
\label{eqeps1}
\end{equation}
where $A_1(R)$ is the morphological lopsidedness and $R_{\rm w}$ is the Gaussian scale length. 

The corresponding relations for $m=2$ and $m=3$ are given by
\begin{equation}
\epsilon_2=\frac{A_2(R)}{\left(\frac{R}{R_{\rm w}}\right)^2+1},
\label{eqeps2}
\end{equation}
and
\begin{equation}
\epsilon_3=\frac{A_3(R)}{(2/7)\left(\frac{R}{R_{\rm w}}\right)^2+1},
\label{eqeps3}
\end{equation}
\section{Results}
\subsection{Morphological lopsidedness}
\label{SectResmorpho}
\begin{figure*}
 \centering
\includegraphics[width=.74\textwidth,viewport=78 93 450 753, clip=]{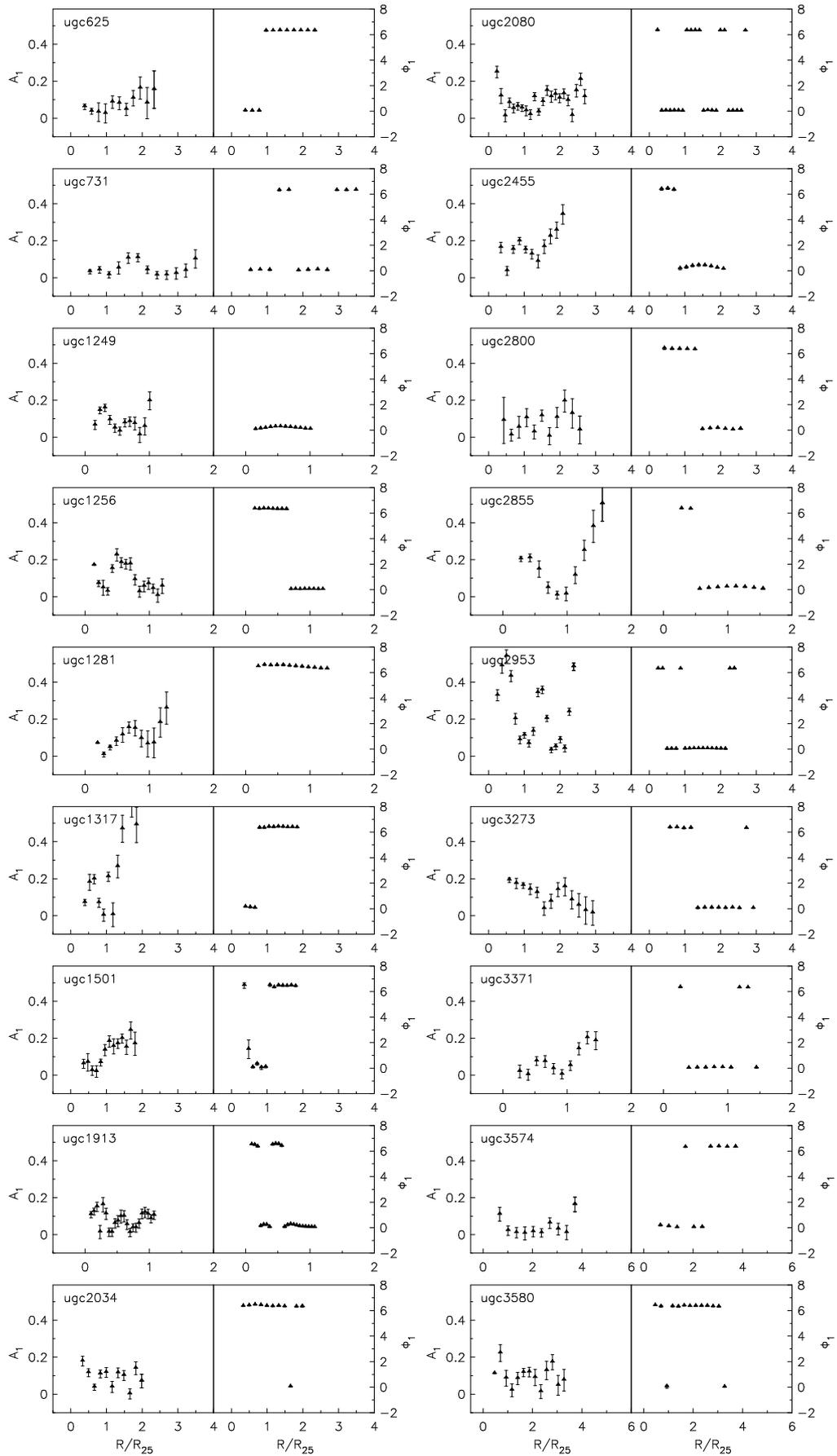}
\caption[]{{\bf Left panels:} morphological lopsidedness $A_1$ as a function of $R/R_{25}$; {\bf right panels:} phase $\phi_1$ as a function of $R/R_{25}$.}
\label{figlop_1}
\end{figure*}

\begin{figure*}
 \centering
\includegraphics[width=.74\textwidth,viewport=78 93 450 753, clip=]{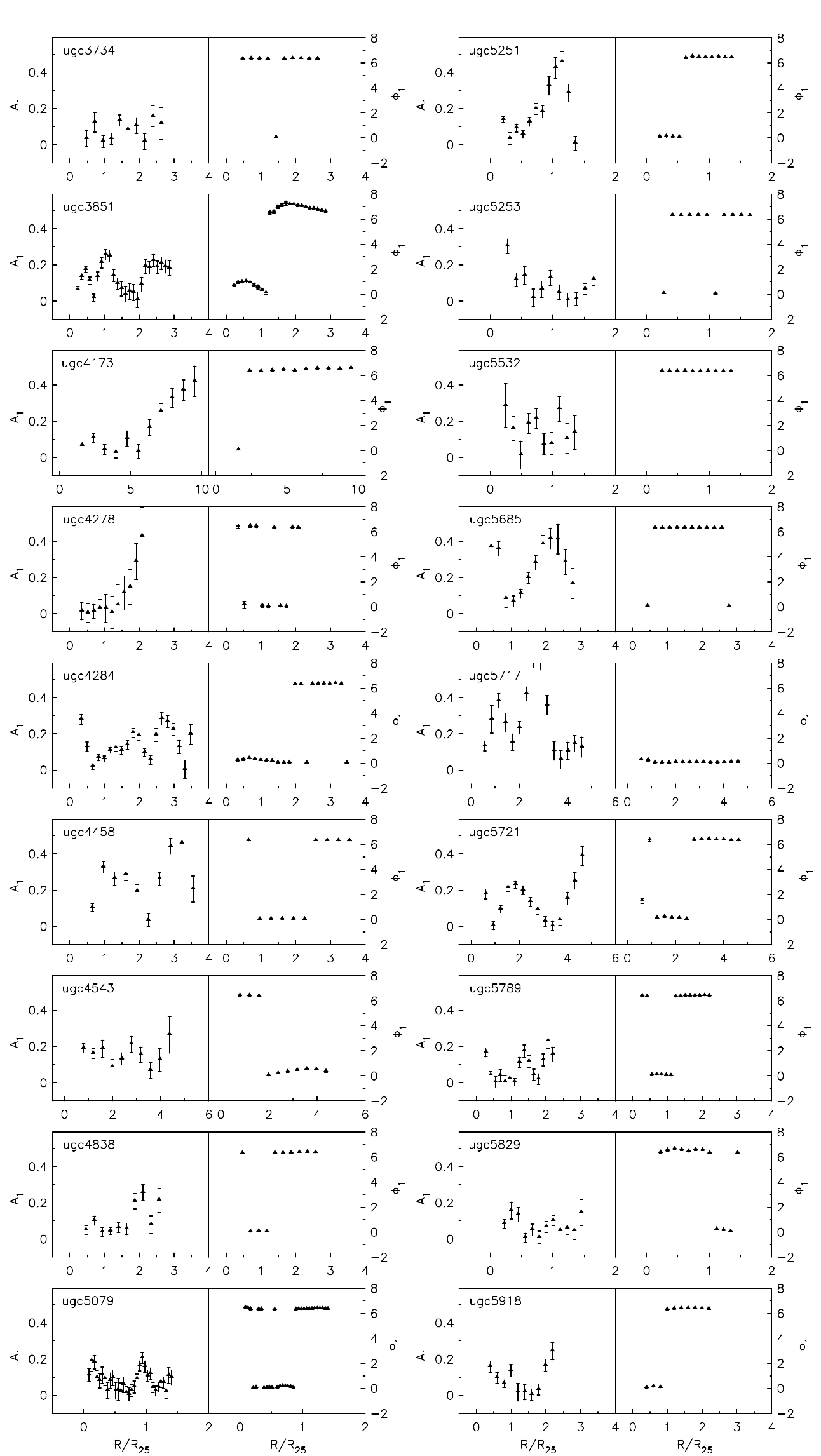}
\caption[]{Fig.~\ref{figlop_1} to be continued.}
\label{figlop_2}
\end{figure*}

\begin{figure*}
 \centering
\includegraphics[width=.74\textwidth,viewport=78 93 450 753, clip=]{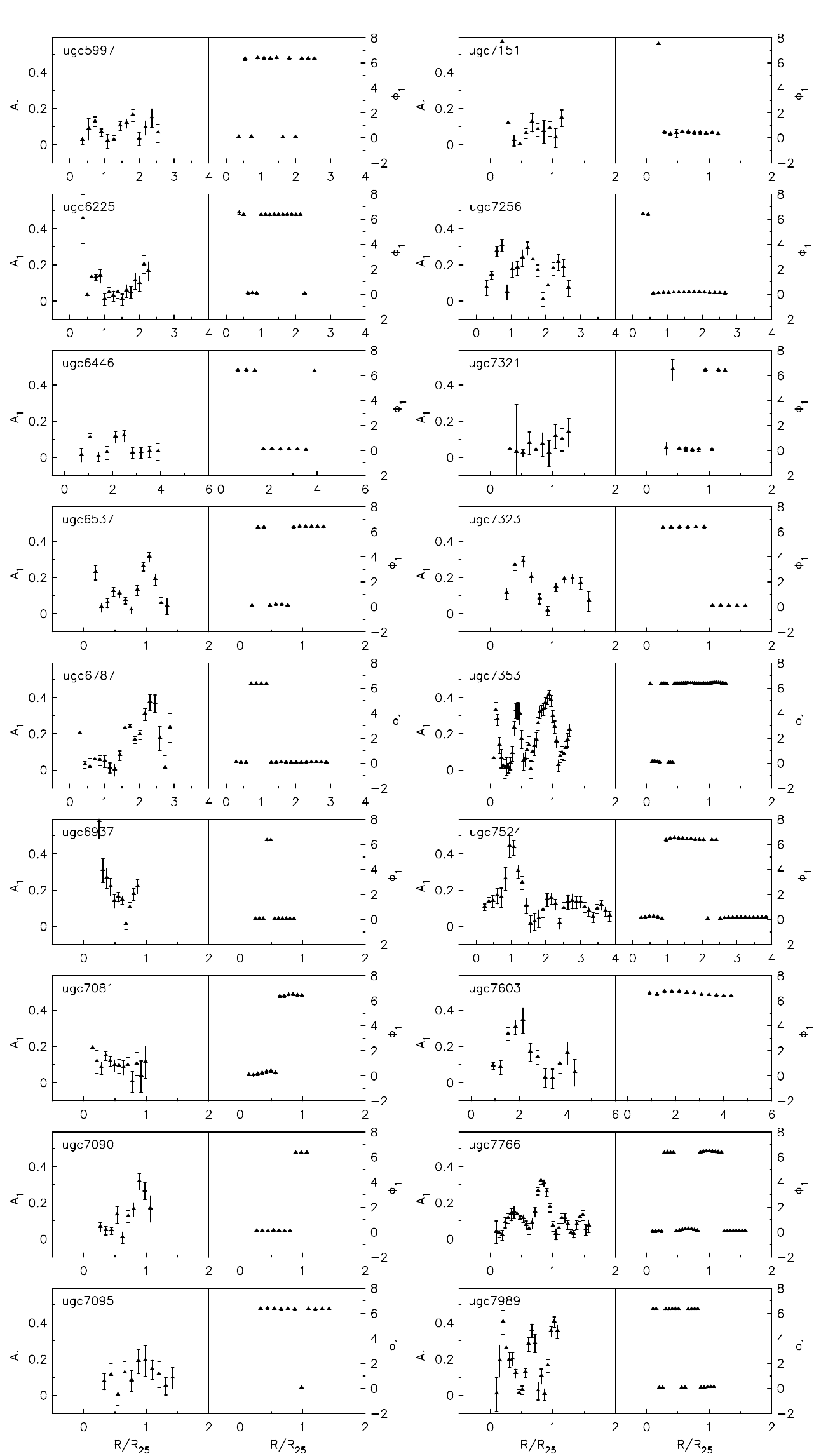}
\caption[]{Fig.~\ref{figlop_1} to be continued.}
\label{figlop_3}
\end{figure*}

\begin{figure*}
 \centering
\includegraphics[width=.74\textwidth,viewport=78 93 450 753, clip=]{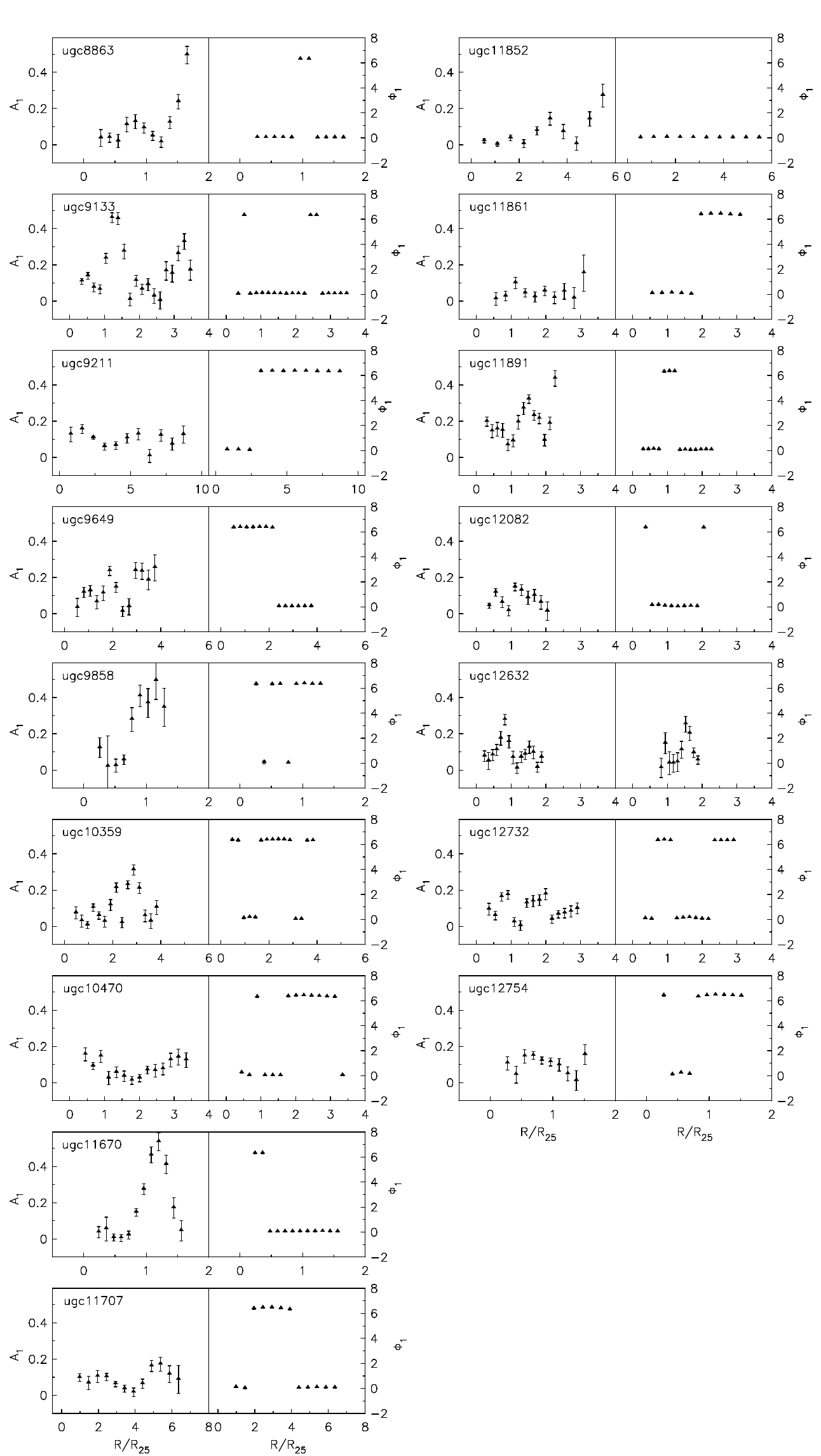}
\caption[]{Fig.~\ref{figlop_1} to be continued.}
\label{figlop_4}
\end{figure*}

Figures~\ref{figlop_1} to \ref{figlop_4} show plots of the amplitude $A_1(R)$ and the phase $\phi_1(R)$ of lopsidedness \emph{vs.} $R/R_{25}$, where $R_{25}$ is the apparent radius of the optical disc. Most galaxies in our sample have \HI\ data extending beyond $R_{25}$ (see Fig. 1, Paper I). Therefore, the Fourier analysis is done beyond the optical radius and in some cases to several times this distance. The optical disc size, $R_{25}$ is typically equal to four to five times the exponential disc scale length \citep{vanderKruit1982}. Thus, in the present paper, the $A_1$ values denoting the lopsided amplitude are measured to four or more disc scale lengths. This is much larger than the maximum distance of 2.5 disc scale lengths to which lopsidedness can be studied in the near-IR \citep[][]{Rix1995,Bournaud2005}, where this limit is set by the high sky background.

The previous studies showed that in the region inside of the optical disc, $A_1$  increases with radius \citep[][]{Rix1995,Saha2007}. This trend was also seen in the \HI\ studies of the Eridanus group galaxies that covered about four disc scale lengths or the optical radius \citep{Angiras2006}. In contrast, in the present work we find that only about 20\%\ of our sample galaxies show a steady increase in the morphological lopsidedness towards large radii. Good examples are UGC\,2455, UGC\,4173 or UGC\,4278. There are also examples of galaxies where the values of $A_1(R)$ scatter significantly with radius (e.g., UGC\,2953 or UGC\,7989). In other galaxies, we discovered pronounced features as, e.g., in UGC\,3851 or UGC\,7353. And finally, we also detected galaxies with a constant value of $A_1(R)$ (e.g., UGC\,3574 or UGC\,7081).

We consider $R_{25}$ to be a limiting radius between the inner and the outer disc. Therefore, we calculated a mean value of the amplitude $A_1(R)$ over all radii smaller than $R_{25}$ and over all radii larger than $R_{25}$. The results are given in Table~\ref{Lopprop}, columns (4) and (5). The last line shows the mean of the averaged amplitudes over all sample galaxies. The mean of $<A_1>$ for large radii is about 0.15, for small radii about 0.11, which means that the inner parts of our sample galaxies are on average slightly less perturbed than the outer parts. All sample galaxies showing amplitudes above these values are considered as lopsided. For large radii, we found 30 galaxies with strong signs of lopsidedness, 38 galaxies show values below 0.15 (for two galaxies we only have data within $R_{25}$). For small radii, about 50\%\ of all sample galaxies are strongly lopsided. 

Note that the typical lopsided amplitude $A_1$ within the optical disc ($R<R_{25}$) of about 0.11 agrees with the value obtained by \citet{Bournaud2005} for the inner half of the optical disc (1.5-2.5 optical disc scale lengths) from near-IR studies of the OSU sample. \citet{Angiras2006} studied the \HI\ asymmetry in 18 galaxies of the Eridanus group and obtained a mean larger than 0.2 in this radial range. Therefore, our WHISP sample shows less lopsidedness than a group case.

As we already discussed in Paper\,I, the existence of warps might affect the determination of the lopsidedness, especially of low to moderately inclined galaxies. However, this effect is not likely to be more pronounced at large radii. This is because the amplitudes of both these features increase with radius. Furthermore, the amplitude of lopsidedness appears to saturate at radii beyond the optical disc.
%
\begin{table*}
\caption{\label{Lopprop}Lopsidedness parameters.}
$$
\begin{tabular}{p{1.0cm}p{1.7cm}p{1.5cm}p{1.5cm}p{1.5cm}p{1.5cm}p{1.5cm}p{1.5cm}p{1.5cm}p{1.0cm}}
\hline\hline
\noalign{\smallskip}
UGC & Hubble Type & $R_{25}$ & $<A_1>$ & $<A_1>$ & $<A_2>$ & $<A_2>$ & $<A_3>$ & $<A_3>$  & $T_{\rm p}$\\
& & [kpc] & ($> 1 R/R_{25}$) & ($< 1 R/R_{25}$) & ($> 1 R/R_{25}$) & ($< 1 R/R_{25}$) & ($> 1 R/R_{25}$) & ($< 1 R/R_{25}$) & \\
(1) & (2) & (3) & (4) & (5) & (6) & (7) & (8) & (9) & (10)\\
\hline
\noalign{\smallskip}
625 & 4 & 13.94 & 0.103 & 0.039 & 0.148 & 0.099 & 0.096 & 0.041 & $-$5.88\\
731 & 9.9 & 2.17 & 0.052 & 0.037 & 0.090 & 0.048 & 0.078 & 0.040 & ...\\
1249 & 8.9 & 7.05 & 0.197 & 0.077 & 0.099 & 0.112 & 0.061 & 0.139 & $-$2.81\\
1256 & 6 & 7.42 & 0.036 & 0.114 & 0.086 & 0.105 & 0.017 & 0.049 & $-$2.68\\
1281 & 7.5 & 4.10 & 0.171 & 0.088 & 0.779 & 0.434 & 0.117 & 0.077 & ...\\
1317 & 4.9 & 23.33 & 0.425 & 0.104 & 0.140 & 0.080 & 0.150 & 0.103 & ...\\
1501 & 7.8 & 3.46 & 0.181 & 0.063 & 0.176 & 0.308 & 0.174 & 0.065 & ...\\
1913 & 7 & 14.52 & 0.096 & 0.079 & 0.107 & 0.214 & 0.110 & 0.065 & $-$5.46\\
2034 & 9.8 & 4.43 & 0.078 & 0.112 & 0.214 & 0.080 & 0.056 & 0.030 & $-$6.80\\
2080 & 6 & 8.50 & 0.101 & 0.091 & 0.134 & 0.047 & 0.060 & 0.033 & ...\\
2455 & 9.9 & 3.27 & 0.194 & 0.138 & 0.223 & 0.045 & 0.121 & 0.047 & ...\\
2800 & 10 & 7.02 & 0.091 & 0.052 & 0.103 & 0.283 & 0.128 & 0.158 & ...\\
2855 & 5 & 9.03 & 0.312 & 0.106 & 0.078 & 0.185 & 0.13 & 0.037 & ...\\
2953 & 3.4 & 8.74 & 0.180 & 0.347 & 0.179 & 0.181 & 0.077 & 0.054 & ...\\
3273 & 9 & 4.56 & 0.087 & 0.177 & 0.128 & 0.101 & 0.090 & 0.038 & ...\\	
3371 & 9.9 & 7.08 & 0.146 & 0.036 & 0.076 & 0.031 & 0.131 & 0.106 & ...\\
3574 & 5.8 & 4.69 & 0.037 & 0.110 & 0.069 & 0.175 & 0.066 & 0.024 & $-$8.70\\
3580 & 1.1 & 5.97 & 0.087 & 0.140 & 0.099 & 0.170 & 0.062 & 0.249 & ...\\
3734 & 4.4 & 4.83 & 0.092 & 0.059 & 0.103 & 0.018 & 0.100 & 0.070 & ...\\
3851 & 9.8 & 2.16 & 0.142 & 0.121 & 0.287 & 0.049 & 0.207 & 0.115 & ...\\
4173 & 9.9 & 1.54 & 0.175 & ... & 0.281 & ... & 0.158 & ...  & ...\\
4278 & 6.4 & 4.40 & 0.151 & 0.016 & 1.213 & 0.550 & 0.245 & 0.027 & $-$4.99\\
4284 & 6 & 4.30 & 0.155 & 0.111 & 0.108 & 0.083 & 0.076 & 0.055 & $-$7.05\\
4458 & 1 & 14.46 & 0.268 & 0.215 & 0.237 & 0.098 & 0.183 & 0.127 & ...\\
4543 & 7.9 & 5.55 & 0.154 & 0.191 & 0.218 & 0.055 & 0.043 & 0.008 & ...\\
4838 & 5.2 & 11.59 & 0.131 & 0.062 & 0.100 & 0.066 & 0.142 & 0.052 & ...\\
5079 & 4 & 15.25 & 0.071 & 0.087 & 0.087 & 0.268 & 0.070 & 0.093 & ...\\
5251 & 4.3 & 14.97 & 0.294 & 0.144 & 0.524 & 0.397 & 0.164 & 0.084 & $-4$.90\\
5253 & 2.4 & 11.14 & 0.051 & 0.129 & 0.043 & 0.095 & 0.120 & 0.097 & $-$5.21\\	
5532 & 3.9 & 24.35 & 0.169 & 0.144 & 0.072 & 0.217 & 0.074 & 0.085 & ...\\
5685 & 4 & 6.78 & 0.258 & 0.271 & 0.133 & 0.134 & 0.043 & 0.026 & $-$6.74\\
5717 & 3.7 & 6.49 & 0.275 & 0.206 & 0.145 & 0.114 & 0.102 & 0.017 & $-$7.41\\
5721 & 6.6 & 1.58 & 0.152 & 0.091 & 0.081 & 0.115 & 0.105 & 0.098 & ...\\
5789 & 6 & 7.44 & 0.109 & 0.045 & 0.085 & 0.090 & 0.037 & 0.061 & ...\\
5829 & 9.8 & 5.85 & 0.079 & 0.072 & 0.061 & 0.115 & 0.131 & 0.041 & ...\\
5918 & 10 & 2.81 & 0.080 & 0.113 & 0.290 & 0.075 & 0.042 & 0.037 & ...\\
5997 & 4 & 8.17 & 0.084 & 0.075 & 0.083 & 0.019 & 0.103 & 0.050 & ...\\
6225 & 6 & 7.18 & 0.074 & 0.176 & 0.108 & 0.077 & 0.048 & 0.076 & $-$7.92\\
6446 & 6.6 & 2.46 & 0.052 & 0.011 & 0.064 & 0.048 & 0.039 & 0.100 & ...\\
6537 & 5.1 & 10.92 & 0.149 & 0.114 & 0.081 & 0.072 & 0.080 & 0.068 & $-$5.71\\
6787 & 1.7 & 9.53 & 0.173 & 0.070 & 0.139 & 0.110 & 0.056 & 0.079 & $-$5.50\\		
6937 & 4 & 20.24 & ... & 0.210 & ... & 0.045 & ... & 0.060 & $-$3.47\\	
7081 & 4.7 & 13.60 & ... & 0.097 & ... & 0.073 & ... & 0.094 & $-$3.32\\
7090 & 5.3 & 8.67 & 0.166 & 0.129 & 0.342 & 0.298 & 0.054 & 0.055 & $-$6.65\\
7095 & 4.1 & 11.57 & 0.099 & 0.109 & 0.106 & 0.142 & 0.158 & 0.065 & $-$6.86\\
7151 & 6 & 2.67 & 0.091 & 0.125 & 0.061 & 0.204 & 0.047 & 0.062 & $-$6.32\\	
7256 & $-$2.7 & 8.33 & 0.167 & 0.168 & 0.267 & 0.255 & 0.186 & 0.041 & ...\\
7321 & 6.6 & 5.01 & 0.116 & 0.043 & 0.738 & 0.495 & 0.084 & 0.042 & ...\\
7323 & 7.9 & 4.48 & 0.152 & 0.160 & 0.133 & 0.065 & 0.202 & 0.045 & $-$5.67\\
7353 & 4 & 19.46 & 0.146 & 0.185 & 0.123 & 0.244 & 0.183 & 0.060 & $-$2.40\\
7524 & 8.9 & 2.12 & 0.121 & 0.182 & 0.109 & 0.175 & 0.093 & 0.041 & $-$8.00\\	
7603 & 7 & 1.60 & 0.152 & 0.092 & 0.082 & 0.061 & 0.046 & 0.061 & ...\\
7766 & 6 & 19.84 & 0.073 & 0.137 & 0.0551 & 0.047 & 0.118 & 0.082 & $-$6.00\\
7989 & 2.2 & 25.91 & 0.377 & 0.171 & 0.079 & 0.254 & 0.113 & 0.121 & $-$3.85\\
8863 & 1 & 14.36 & 0.184 & 0.072 & 0.279 & 0.355 & 0.113 & 0.102 & $-$8.15\\
9133 & 2.4 & 22.77 & 0.188 & 0.097 & 0.070 & 0.064 & 0.073 & 0.012 & ...\\
9211 & 9.9 & 1.16 & 0.095 & 0.127 & 0.073 & 0.060 & 0.067 & 0.008 & ...\\
9649 & 3 & 2.08 & 0.150 & 0.076 & 0.066 & 0.021 & 0.111 & 0.011 & ...\\
9858 & 4 & 21.61 & 0.403 & 0.152 & 0.304 & 0.157 & 0.226 & 0.079 & $-$6.54\\
\end{tabular}
$$
\end{table*}
\begin{table*}
$$
\begin{tabular}{p{1.0cm}p{1.7cm}p{1.5cm}p{1.5cm}p{1.5cm}p{1.5cm}p{1.5cm}p{1.5cm}p{1.5cm}p{1.0cm}}
10359 & 5.6 & 4.86 & 0.124 & 0.037 & 0.159 & 0.051 & 0.126 & 0.033 & ...\\
10470 & 4 & 6.90 & 0.069 & 0.131 & 0.230 & 0.038 & 0.124 & 0.066 & ...\\
11670 & 0.5 & 7.70 & 0.325 & 0.078 & 0.104 & 0.231 & 0.172 & 0.065 & ...\\
11707 & 8 & 2.37 & 0.090 & 0.097 & 0.122 & 0.033 & 0.052 & 0.057 & ...\\
11852 & 1 & 10.61 & 0.085 & 0.021 & 0.262 & 0.023 & 0.039 & 0.015 & ...\\
11861 & 7.8 & 6.49 & 0.058 & 0.020 & 0.145 & 0.121 & 0.052 & 0.039 & ...\\
11891 & 9.9 & 4.33 & 0.227 & 0.143 & 0.069 & 0.077 & 0.077 & 0.043 & ...\\
12082 & 8.8 & 3.95 & 0.091 & 0.060 & 0.230 & 0.065 & 0.105 & 0.035 & $-$6.71\\
12632 & 8.7 & 4.28 & 0.068 & 0.133 & 0.050 & 0.056 & 0.062 & 0.079 & $-$6.24\\
12732 & 8.7 & 5.29 & 0.089 & 0.123 & 0.105 & 0.107 & 0.097 & 0.061 & $-$6.35\\
12754 & 6 & 4.70 & 0.078 & 0.114 & 0.047 & 0.118 & 0.152 & 0.057 & $-$5.99\\
\hline
\noalign{\smallskip}
mean & ... & ... & 0.148 & 0.111 & 0.177 & 0.136 & 0.105 & 0.063 & ...\\
\hline	
\end{tabular}
$$
\footnotesize{ 
Notes: (1) galaxy name from the UGC catalogue; (2) morphological type following the classification by \citet{deVaucouleurs1979}; (3) apparent radius from the HyperLeda database; (4) to (9) the mean values of the parameters for morphological lopsidedness as obtained in this paper; (10) tidal parameter as derived from Eq.~\ref{eqtidalparam}.}
\end{table*}

A striking result from Figs.~\ref{figlop_1} to \ref{figlop_4} is that the phase $\phi_1(R)$ is nearly constant with radius, which indicates that $m=1$ is a global mode.  Note that $\phi_1(R)$ is given in radians. The apparently different values are the same angle modulo 2$\phi$. This result will be discussed in detail in Sect.~\ref{SectPhase}.
\subsection{A comparison of $<A_1>$, $<A_2>$, and $<A_3>$}
\begin{figure}
\centering
\includegraphics[width=.5\textwidth,viewport=78 393 370 683, clip=]{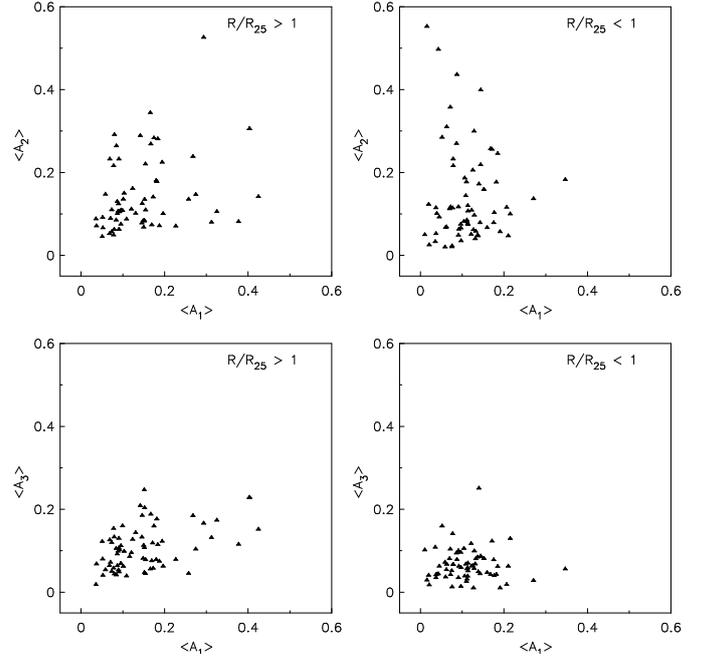}
\caption[]{A comparison of the mean values of the fractional amplitudes of the first three Fourier components of the \HI\ surface density distribution. {\bf Upper row}: $<A_2>$ \emph{vs.} $<A_1>$ for large and small radii, beyond and within the optical radius respectively. The $<A_2>$ values denoting bars and two-armed spirals dominate over lopsidedness within the optical radius, for amplitudes $> 0.2$. Thus, strong bars are more likely in the inner galaxy than very strong lopsidedness. {\bf Lower row}: $<A_3>$ \emph{vs.} $<A_1>$ for large and small radii, the $<A_3>$ values are smaller than lopsided amplitudes, especially at large radii.}
\label{figcompa}
\end{figure}
We now want to compare the mean normalised amplitudes averaged over $R/R_{25}>1$ and $R/R_{25}<1$ for the first three Fourier components of the \HI\ surface density distribution. The upper two panels in Fig.~\ref{figcompa} show $<A_2>$ \emph{vs.} $<A_1>$ averaged over large and small radii. We see a comparable distribution for large radii, whereas at small radii, the $m=2$ perturbations are much stronger. As mentioned in Sect.~\ref{secmorpholop}, $A_2$ denotes perturbations caused by a bar or spiral arms. Although spiral arms can be found at all radii in a disc, bars will dominate in the central parts. This shows that the amplitudes of the $m=2$ mode are comparable with the $m=1$ mode at large radii, but clearly dominate at small radii. $<A_3>$ is generally smaller than $<A_1>$ independent of the radial range we look at (Fig.~\ref{figcompa}, lower panels).

We calculated the mean of $A_1$, $A_2$, and $A_3$ for the two different radial ranges: confirming the results from the graphs, the mean of $A_2$ is lower than the mean of $A_1$ for all radii. The mean of $A_3$ is lower for all radii (see Table~\ref{Lopprop}). 
\subsection{Lopsided potential}
As shown in Sect.~\ref{SectPerturbedpot}, we can derive the perturbation of the halo up to third order from the parameters for the morphological lopsidedness. The values are given in the last three columns of Table~\ref{Lopproprw}. For completeness reasons we also list the values for the fractional Fourier amplitudes $A_1$, $A_2$, and $A_3$ between 1 and 2 Gaussian scale lengths in Table~\ref{Lopproprw}. We calculated the mean of all parameters. In agreement with the results for the morphological lopsidedness (see Table~\ref{Lopproprw}, columns 2 to 4), the mean of $<\epsilon_{2}>$ is higher than the mean of $<\epsilon_{1}>$, whereas the mean of $<\epsilon_{3}>$ slightly lower than the mean of $<\epsilon_{2}>$.
\section{Discussion}
\subsection{Variation in lopsidedness with radius}
\label{Sectlopatlargerad}
As mentioned in Sect.~\ref{SectResmorpho}, a pronounced increase in the morphological lopsidedness is only found in about 20\%\ of all sample galaxies. This is somewhat surprising as near-IR observations at smaller radii show that an increase in $A_1$ with radius seems to be a common feature in disc galaxies \citep{Rix1995}. Here, the neutral gas can be traced out to several times the apparent optical radius $R_{25}$. At these large radii, the increase in $A_1$ does not continue, it rather appears to saturate. The origin of this difference is not understood and presents an important clue for the origin of the disc lopsidedness at large radii. Unfortunately, a direct comparison of the behaviour of the stellar and gaseous component cannot be done here since the inner parts of the galaxies, on which scales \citet{Rix1995} found an increase of lopsidedness, are not resolved in our \HI\ maps.

\begin{figure*}
 \centering
\includegraphics[width=.99\textwidth,viewport=48 403 575 663, clip=]{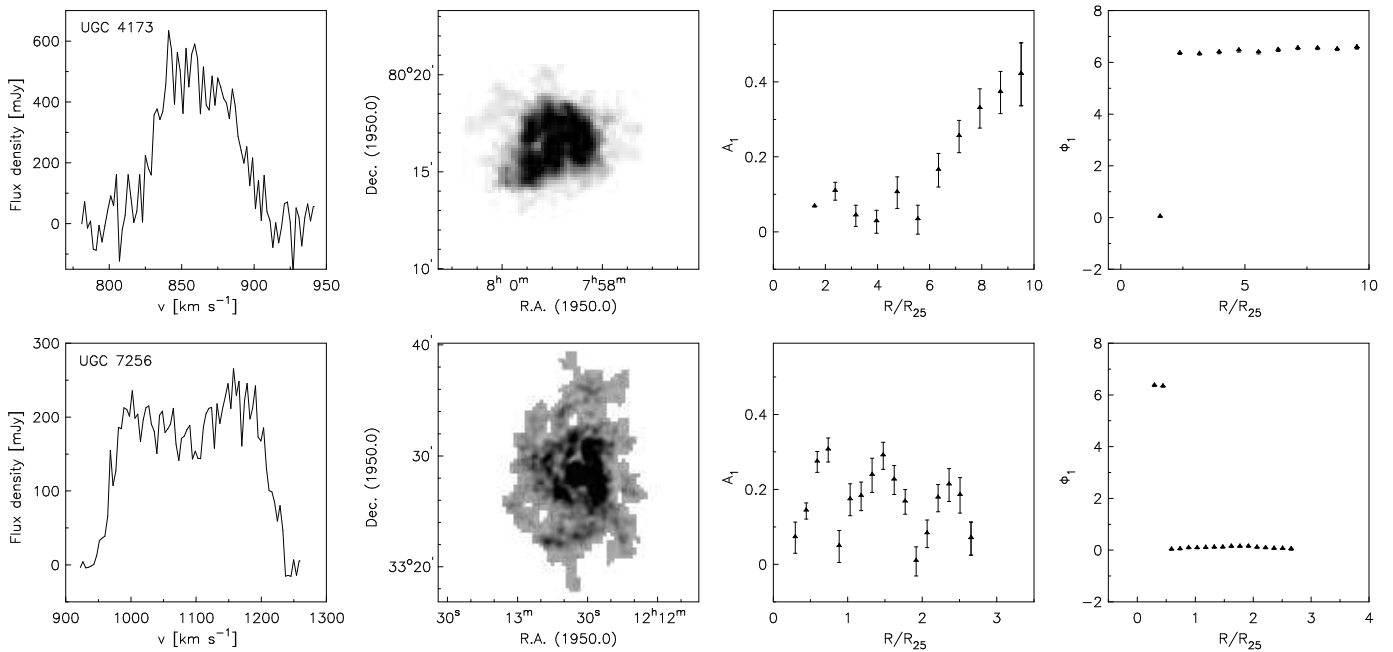}
\caption[]{Global profile (left panels), \HI\ intensity distribution and the distribution of $A_1$ (middle panels), and the distribution of the phase $\phi_1$ (right panels) of two example galaxies, UGC\,4173 (upper row) and UGC\,7256 (lower row).}
\label{figexamples}
\end{figure*}

To get a clearer picture, we selected two galaxies and compared their global profile, the \HI\ morphology, and the distributions of $A_1$  and $\phi_1$ with radius. UGC\,4173 (see Fig.~\ref{figexamples}, upper row) represents a type of galaxy where we found the expected behaviour of increasing lopsidedness with radius. As can be seen, the \HI\ distribution is quite smooth with a clear asymmetry in the south-east. UGC\,7256 (Fig.~\ref{figexamples}, lower row) is an example of a galaxy where we found strong fluctuations, so-called wiggles, in the distribution of $A_1$. The \HI\ intensity map shows a spiral galaxy with pronounced arms. The global profile has the typical two maxima. However, the intensities of the peaks slightly differ from each other, indicating lopsidedness. Furthermore, in both cases the phase $\phi_1 (R)$ is nearly constant with radius indicating  a global $m=1$ mode. We therefore suggest that the fluctuations in the distribution of $A_1$ reflect the local \HI\ morphology (i.e., local spiral arms) superposed on top of the global lopsidedness. 

A natural way to explain the origin of these local spiral features would be via the swing amplification \citep[][]{Goldreich1965,Toomre1981}. Such fluctuations are not apparent in the near-IR studies that measure the asymmetry in the underlying old stellar distribution \citep[see, e.g., Fig.~1 in][]{Rix1995}. This could be explained by the fact that a low dispersion component like gas has a stronger amplitude for the swing amplification feature than stars \citep{Jog1992}. This is true even in the linear regime studied in the above swing amplification papers, and in a real galaxy the gas will be more likely to exhibit a further non-linear growth in the amplitude.

We did a cross-check to see if this overall picture is reasonable. If the galaxy rotation curve shows a linear increase with radius, i.e., when it denotes a region of solid body rotation so that the angular speed is constant, the swing amplification idea cannot work. Hence, the net $A_1$ is then likely to be a pure underlying global $m=1$ component showing a radial increase in the lopsided amplitude. This is exactly confirmed for galaxies like UGC\,4173 or UGC\,4278 which show a smooth radial increase in $A_1$ (Fig.~\ref{figlop_2}, this paper) and both have solid body rotation in the corresponding radial region as seen in Fig.~A1, Paper\,I. In contrast, in the region of the flat rotation curve, the differential rotation is significant and the swing amplification is effective, resulting in the fluctuations -- as confirmed for UGC 731 or UGC 2080 which show a flat rotation (Fig.~A1, Paper I) and have a fluctuating A$_1$ in the same radial region (Fig.~\ref{figlop_1}, this paper). Since flat rotation curves are common, this explains why most of our sample galaxies show fluctuations in the amplitude for lopsidedness $A_1$ in \HI.

An important point to note is that while the lopsided amplitude $A_1$ does show large, local fluctuations, its overall value increases with radius, linearly up to the optical radius and then somewhat slower. Thus, we can  explain why lopsidedness was first noted in \HI\ in the outer parts of galaxies like M\,101 \citep{Baldwin1980}. Therefore, \HI\ still serves as an excellent tracer of lopsidedness in the outer galactic discs.

\begin{figure}
 \centering
\includegraphics[width=.4\textwidth,viewport=68 518 260 713, clip=]{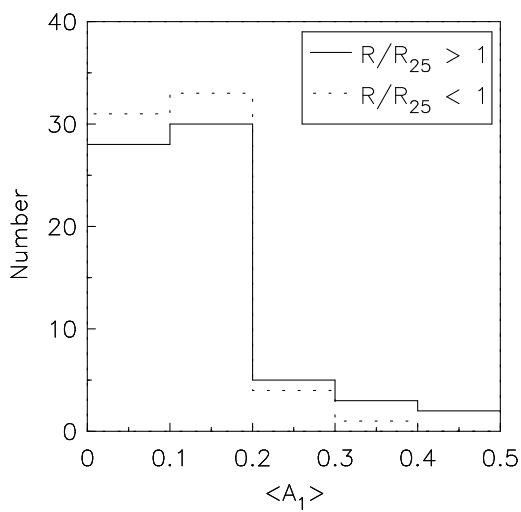}
\caption[]{The distribution of $<A_1>$ for small (dotted histogram) and large radii (solid histogram).}
\label{figa1distribution}
\end{figure}

Figure~\ref{figa1distribution} shows the distribution of $<A_1>$ for small (dotted histogram) and large radii (solid histogram). Surprisingly, both distributions are comparable: most galaxies have a lopsidedness parameter below 0.2, only a few have higher values. Thus, while the lopsidedness is lower in the inner parts than beyond the optical disc (Table~\ref{Lopprop}), the increase is not strong and the $A_1$ values seem to saturate beyond the optical radius. This needs to be studied by future theoretical work.
\subsection{The phase $\phi_1(R)$}
\label{SectPhase}
As already mentioned in Sect.~\ref{SectResmorpho}, the phase $\phi_1(R)$ is in most cases nearly constant with radius. This trend was already noted from smaller samples studied to a radial extent less than the optical radius \citep[][]{Rix1995,Angiras2007}. Remarkably, the phase remains nearly constant even when the lopsided amplitude shows a wide variation with radius as discussed above. This indicates that $m=1$ is a global mode and denotes a perturbation in the underlying  mass distribution. Since the stars typically contribute less than the gas to the surface density beyond the optical radius or about four to five disc scale lengths as in our Galaxy \citep{Narayan2002}, the constant phase most likely denotes the origin of the lopsided potential to be the distorted dark matter halo, to which the \HI\ in the outer disc responds. This confirms our assumption made in deriving the perturbation potential (Sect.~\ref{SectPerturbedpot}). It also makes it harder to support gas accretion as a mechanism for lopsidedness which was proposed by \citet{Bournaud2005}.
\subsection{Morphological vs. kinematic lopsidedness}
\label{Sectmorphovskin}
It has been argued that kinematic and morphological asymmetries are causally connected \citep{Jog2002}. However, \citet{Kornreich2002} showed that both asymmetries are not always correlated. We therefore want to find out how kinematic and morphological lopsidedness are correlated in the WHISP sample.

\begin{figure}
 \centering
\includegraphics[width=.45\textwidth,viewport=78 535 240 673, clip=]{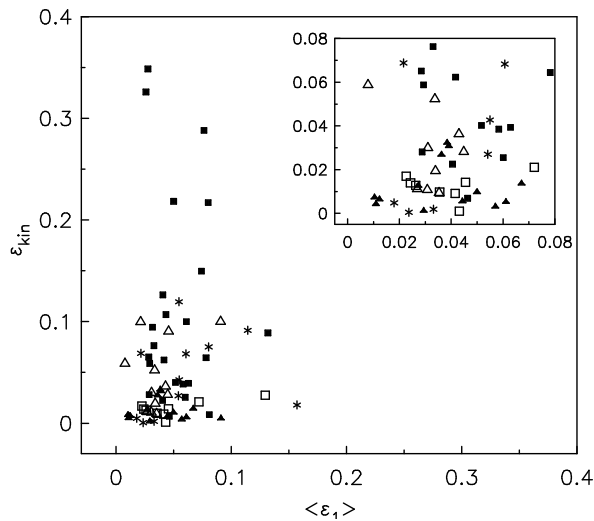}
\caption[]{The perturbation parameter in the lopsided potential obtained by the analysis of the kinematic data $<\epsilon_{\rm kin}>$ \emph {vs.} the one obtained from the morphological data $<\epsilon_{1}>$. The symbols represent the five kinematic classes as defined in Paper I. Type 1: receding/approaching sides agree on all scales (solid triangles); Type 2: constant offset of receding/approaching sides (solid squares); Type 3: differences only at large radii (open triangles); Type 4: differences only at small radii (open squares); Type 5: receding and approaching sides change sides (stars).}
\label{figepsvsa1}
\end{figure}

A comparison of $\epsilon_{\rm kin}$ and $<\epsilon_1>$, the perturbation parameter in the lopsided potential obtained by analysing the kinematic and morphological data respectively, is presented in Fig.~\ref{figepsvsa1}. Following the classification in Paper I, we distinguish between five kinematic classes:
\begin{itemize}
\item \emph{Type 1}: receding and approaching sides agree on all scales (solid triangles),
\item \emph{Type 2}: constant offset of receding and approaching sides (solid squares),
\item \emph{Type 3}: differences between receding and approaching sides only at large radii (open triangles),
\item \emph{Type 4}: differences between receding and approaching sides only at small radii (open squares), 
\item \emph{Type 5}: receding and approaching sides change sides (stars).
\end{itemize}
In general, it can be seen that the perturbation parameters as calculated from the morphological and kinematic data are comparable. There are a few discrepancies where galaxies show a much higher value for $\epsilon_{\rm kin}$ than for $\epsilon_1$ (UGC\,1249, UGC\,2034, UGC\,2080, UGC\,4458, UGC\,4543). These galaxies are all classified as Type~2. In case of UGC\,2034, the rotation velocity could not be well defined as the rotation curve shows a second rise at higher radii. We took an average value of the first plateau as a lower limit for $v_{\rm rot}$, which results in an upper limit for $\epsilon_{\rm kin}$. In case of UGC\,2080, we did not find any local asymmetries in the velocity field, but a global offset of the iso-velocity contours (see Paper\,I, Sect. 4.1). This results in a high $\epsilon_{kin}$, while the galaxy is only moderately lopsided morphologically. Thus, the way the lopsidedness is measured may produce a spurious discrepancy between kinematic and morphological lopsidedness in some galaxies. The same seems to be true for the other three galaxies.

There are also a few cases where the the perturbation parameter obtained from the morphological data is higher than the one obtained from the kinematic data. A good example is UGC\,8863 (Paper\,I, Sect.~4.1) where the perturbation parameter obtained from the kinematic data is feeble as the discrepancies of receding and approaching sides are small in the outer parts where the maximum rotation velocities have been measured. The perturbation parameter obtained from the morphological data, however, is much higher.

We can conclude that morphological and kinematic lopsidedness are generally comparable, which is in good agreement with \citet{Jog2002}.
\subsection{Tidal origin of lopsidedness}
As mentioned in the introduction, the physical origin of lopsidedness and its life time are closely connected, i.e., different mechanisms lead to different life times of lopsidedness. We here study the correlations of $A_1$ with galaxy type and the strength of tidal interaction. As $A_1$ has been measured to very large distances, we will be able to better constrain the mechanism of the origin of lopsidedness in comparison to earlier studies.

\begin{figure}
 \centering
\includegraphics[width=.49\textwidth,viewport=78 530 390 683, clip=]{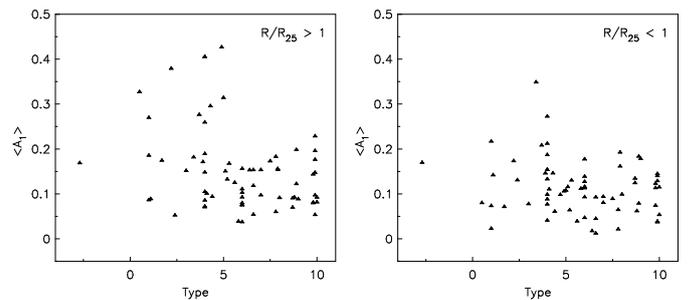}
\caption[]{Morphological lopsidedness \emph{vs.} morphological type for large (left panel) and small (right panel) radii.}
\label{figklopvstype}
\end{figure}

In a first step, we compare the morphological lopsidedness $<A_1>$ averaged over large and small radii (Fig.~\ref{figklopvstype}, left and right panels respectively) to the morphological type. The early-type spiral galaxies seem to be more strongly lopsided, especially in the outer radial regions. Since a tidal interaction drives a galaxy towards early-type, this result of high lopsidedness in early-type galaxies indicates tidal encounters as the mechanism for the origin of lopsidedness \citep[see also][]{Angiras2006}.
\label{SectOrigin}

\begin{figure}
 \centering
\includegraphics[width=.4\textwidth,viewport=78 533 240 673, clip=]{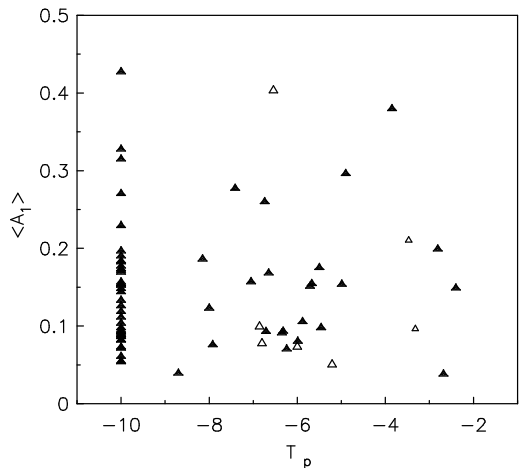}
\caption[]{The amplitude of morphological lopsidedness $<A_1>$ averaged over large radii \emph{vs.} the tidal parameter $T_{\rm p}$ (large filled triangles). In some cases, up to two companions could not be included in the calculation of $T_{\rm p}$ due to the lack of data. This gives a lower limit for $T_{\rm p}$ (large open triangles). In two cases, the \HI\ emission could only be detected within $R_{25}$ so that $<A_1>$ averaged over small radii is plotted \emph{vs.} $T_{\rm p}$ (small open triangles). There is no clear correlation of the lopsided amplitude with the tidal parameter.}
\label{figtidalparm}
\end{figure}

As a next step, we look at the environment of our sample galaxies. The degree of lopsidedness depends on the distance between the main galaxy and its companions, as well as on the mass ratio of the main galaxy and its companions. We therefore use the so-called tidal parameter to quantify the effect of tidal forces \citep{Bournaud2005}, which is given by
\begin{equation}
T_p=\log \left( \sum_i\frac{M_i}{M_0}\left(\frac{R_0}{D_i}\right)^3\right),
\label{eqtidalparam}
\end{equation}
where $M_i$ is the mass of each companion, $M_0$ is the mass of the main galaxy, $R_0$ is the scale length and $D_i$ is the distance of each companion from the main galaxy. The ratio of the masses is estimated from the ratio of absolute blue magnitudes. We take $R_{25}/4$ as an estimate for the exponential disc scale length (van der Kruit \& Searle 1982). The search for neighbours was done using the Tully Nearby Galaxies Catalogue \citep{Tully1988}. We only include companions within a radius of 2.5\degr\ and with measured radial velocities within 500\skms. Although almost all galaxies in our sample are members of groups or associations, many of them appear to be isolated on the basis of the above mentioned criteria. We find 31 galaxies to have between one and 15 companions. The $T_{\rm p}$ values of these 31 galaxies are plotted in Fig.~\ref{figtidalparm} \emph{vs.} the parameter for morphological lopsidedness $<A_1>$ averaged over large radii (large filled triangles). In a few cases, up to two companions do not have any data for the absolute blue magnitude, which gives us a lower limit for the tidal parameter (large open triangles). In two cases, only data within $R_{25}$ are available so that we show $<A_1>$ for small radii (small open triangles).

In good agreement with \citet{Bournaud2005}, no correlation between $<A_1>$ and $T_{\rm p}$ is found. Strong lopsidedness is as common in more isolated galaxies (smaller $T_{\rm p}$) as in galaxies with many companions. This implies that a different, internal mechanism for the origin of lopsidedness has to take place in isolated galaxies to cause them to be lopsided. Alternatively, if the lopsidedness is long-lived as indicated by theoretical studies \citep{Saha2007}, or if it arises due to satellite accretion as proposed by \citet{Zaritsky1997} and shown by simulations \citep{Bournaud2005}, then these two factors could explain why isolated galaxies could be lopsided as well. If lopsidedness is long-lived, then there would be no correlation of $A_1$ with a tidal parameter even if the origin of lopsidedness is due to a tidal encounter. As has been shown above, the higher lopsidedness seen in early-type galaxies (Fig.~\ref{figklopvstype}) suggests a tidal origin. Thus, these two points together indicate a tidal encounter as the likely origin of lopsidedness in the WHISP sample studied here.

It has to be noted that this analysis is limited by the incompleteness of the Tully Nearby Galaxies Catalogue. In some cases not all relevant parameters were listed in the catalogue so that these companions could not be included in the tidal parameter calculation. Furthermore, it is likely that not all small companions have been detected so far. Still, this method gives a good indicator of the overall local environment of the galaxies, even though some low-mass companions might be missing.
\section{Summary}
We carried out a harmonic analysis of the \HI\ surface density distribution and measured the morphological lopsidedness in a sample of 70 galaxies selected from the WHISP survey. The amplitude of lopsidedness, $A_1$, is defined to be the fractional amplitude of the first azimuthal Fourier component. The most interesting aspect of this study is that due to the high sensitivity data, it was possible to measure lopsidedness for the first time out to large radial distances. Typically, lopsidedness is measured out to one to four optical radii and in a very few cases even up to ten optical radii (Fig. 1, Paper I).

The main results from our study can be summarised as follows:
\begin{itemize}
\item{The mean amplitude of lopsidedness, $<A_1>$ within the optical disc is high, about 0.1. The values of $A_1$ increase radially within this radial range. Beyond this, the values seem to saturate.}
\item{The phase is remarkably constant with radius indicating that the lopsidedness is a global mode, as was argued earlier \citep{Saha2007}.}
\item{In a few cases, the values of $A_1$ increase with radius, but in most cases they show small-scale fluctuations. We argue that the latter are due to local spiral features arising via swing amplification.}
\item{As shown in Paper\,I, the galaxies can be divided into five groups based on their rotation velocity patterns. The most common types are galaxies whose velocity fields are globally distorted (types 2 and 5). In general, kinematic and morphological lopsidedness agree quite well.}
\item{Early-type galaxies seem to be more lopsided than late-type galaxies. We did not find a correlation between lopsidedness and the tidal parameter indicating the strength of interactions for the sample galaxies.}
\end{itemize}
The results from both the morphological and kinematic studies indicate that tidal encounters are the main generating mechanism for the halo lopsidedness, to which the disc responds.
\begin{acknowledgements}
The authors would like to thank the anonymous referee for the constructive feedback which helped to improve this paper.\\
We want to thank Thijs van der Hulst for providing us with the WHISP data cubes before they became publicly available. C.~J. would like to thank DFG (Germany) and INSA (India) for supporting a visit to Germany in October 2007 under INSA-DFG Exchange Programme, during which this collaboration was started. We made use of NASA's Astrophysics Data System (ADS) Bibliographic Services and the NASA/IPAC Extragalactic Database (NED) which is operated by the Jet Propulsion Laboratory, California Institute of Technology, under contract with the National Aeronautics and Space Administration. We acknowledge the usage of the HyperLeda database (http://leda.univ-lyon1.fr).
\end{acknowledgements}

\bibliographystyle{aa}
\bibliography{bibliography}

\begin{appendix}
\section{Lopsided potential}
We here present the perturbation parameters in the lopsided potential as obtained from the morphological data up to the third Fourier component. For a better understanding we also add the Gaussian scale lengths, the parameters for morphological lopsidedness up to third order and the perturbation parameter in the lopsided potential obtained from the kinematic data in Paper~I.
\begin{table*}
\caption{\label{Lopproprw}Lopsidedness parameters.}
$$
\begin{tabular}{p{1.5cm}p{1.5cm}p{1.7cm}p{1.7cm}p{1.7cm}p{1.5cm}p{1.7cm}p{1.7cm}p{1.7cm}}
\hline\hline
\noalign{\smallskip}
UGC & $R_w$ & $<A_1>$ &  $<A_2>$ &  $<A_3>$ & $<\epsilon_{\rm{kin}}>$ & $<\epsilon_{1}>$ & $<\epsilon_{2}>$ & $<\epsilon_{3}>$\\
& [kpc] & ($1<R_w<2$) & ($1<R_w<2$) & ($1<R_w<2$) & & ($1<R_w<2$) & ($1<R_w<2$) & ($1<R_w<2$)\\
(1) & (2) & (3) & (4) & (5) & (6) & (7) & (8) & (9)\\
\hline
\noalign{\smallskip}
625 & 16.35 & 0.106 & 0.157 & 0.107 & 0.065 & 0.029 & 0.052 & 0.058 \\
731 & 3.38 & 0.051 & 0.060 & 0.073 & 0.001 & 0.029 & 0.020 & 0.043\\
1249 & 3.77 & 0.076 & 0.151 & 0.138 & 0.326 & 0.026 & 0.054 & 0.084\\
1256 & 4.68 & 0.078 & 0.084 & 0.056 & 0.005 & 0.044 & 0.030 & 0.037\\
1281 & 2.36 & 0.109 & 0.588 & 0.098 & 0.009 & 0.050 & 0.192 & 0.062\\
1317 & 20.61 & 0.355 & 0.139 & 0.141 & 0.008 & 0.081 & 0.047 & 0.095\\
1501 & 2.62 & 0.152 & 0.291 & 0.126 & 0.091 & 0.046 & 0.100 & 0.070\\
1913 & 6.67 & 0.064 & 0.238 & 0.083 & 0.013 & 0.026 & 0.070 & 0.055\\
2034 & 4.80 & 0.078 & 0.214 & 0.056 &  0.349 & 0.028 & 0.064 & 0.033\\
2080 & 14.01 & 0.119 & 0.134 & 0.051 &  0.218 & 0.050 & 0.042 & 0.034\\
2455 & 2.88 & 0.152 & 0.260 & 0.062 &  0.107 & 0.043 & 0.069 & 0.037\\
2800 & 9.24 & 0.098 & 0.106 & 0.138 &  0.011 & 0.031 & 0.026 & 0.071\\
2855 & 6.93 & 0.154 & 0.123 & 0.099 &  0.019 & 0.034 &  0.047 & 0.061\\
2953 & 17.30 & 0.214 & 0.309 & 0.098 &  0.028 & 0.130 & 0.138 & 0.073\\
3273 & 5.63 & 0.104 & 0.149 & 0.048 &  0.009 & 0.035 & 0.047 & 0.027\\	
3371 & 5.63 & 0.118 & 0.064 & 0.114 &  0.028 & 0.029 & 0.017 & 0.065\\
3574 & 7.91 & 0.025 & 0.061 & 0.061 &  0.059 & 0.008 & 0.021 & 0.036\\
3580 & 9.39 & 0.098 & 0.087 & 0.057 &  0.030 & 0.039 & 0.022 & 0.037 \\
3734 & 6.62 & 0.102 & 0.118 & 0.108 &  0.126 & 0.041 & 0.034 & 0.061\\
3851 & 2.89 & 0.116 & 0.301 & 0.214 &  0.052 & 0.034 & 0.081 & 0.136\\
4173 & 8.01 & 0.262 & 0.432 & 0.225 &  0.150 & 0.074 & 0.128 & 0.138\\
4278 & 3.41 & 0.028 & 1.218 & 0.050 &  0.007 & 0.010 & 0.391 & 0.031\\
4284 & 7.53 & 0.168 & 0.134 & 0.078 &  0.005 & 0.061 & 0.041 & 0.047 \\
4458 & 21.16 & 0.243 & 0.303 & 0.144 &  0.217 & 0.080 & 0.079 & 0.077\\
4543 & 14.73 & 0.164 & 0.307 & 0.065 &  0.288 & 0.076 & 0.103 & 0.041\\
4838 & 15.80 & 0.145 & 0.100 & 0.149 &  0.007 & 0.046 & 0.030 & 0.094\\
5079 & 10.21 & 0.075 & 0.162 & 0.085 &  0.014 & 0.024 & 0.062 & 0.053\\
5251 & 9.19 & 0.286 & 0.612 & 0.161 &  0.100 & 0.091 & 0.201 & 0.096\\
5253 & 15.79 & 0.094 & 0.029 & 0.040 &  0.100 & 0.061 & 0.013 & 0.029 \\	
5532 & 21.52 & 0.146 & 0.056 & 0.086 &  0.039 & 0.063 & 0.018 & 0.058\\
5685 & 10.24 & 0.324 & 0.119 & 0.051 &  0.091 & 0.115 & 0.040 & 0.032\\
5717 & 16.47 & 0.265 & 0.155 & 0.100 &  0.018 & 0.157 & 0.047 & 0.060 \\
5721 & 4.23 & 0.136 & 0.089 & 0.143 &  0.062 & 0.042 & 0.033 & 0.090\\
5789 & 7.55 & 0.085 & 0.067 & 0.047 &  0.012 & 0.027 & 0.025 & 0.032\\
5829 & 3.69 & 0.055 & 0.058 & 0.096 &  0.005 & 0.018 & 0.021 & 0.056\\
5918 & 3.47 & 0.092 & 0.309 & 0.044 &  0.100 & 0.022 & 0.096 & 0.028\\
5997 & 10.79 & 0.102 & 0.092 & 0.125 &  0.026 & 0.036 & 0.025 & 0.075\\
6225 & 8.63 & 0.084 & 0.100 & 0.053 &  0.017 & 0.023 & 0.031 & 0.032\\
6446 & 6.39 & 0.027 & 0.067 & 0.044 &  0.006 & 0.012 & 0.026 & 0.030\\
6537 & 8.09 & 0.143 & 0.077 & 0.080 &  0.027 & 0.054 & 0.027 & 0.052\\
6787 & 12.07 & 0.216 & 0.126 & 0.072 &  0.039 & 0.058 & 0.042 & 0.047\\
6937 & 13.74 & 0.165 & 0.080 & 0.111 &  0.003 & 0.091 & 0.032 & 0.079\\	
7081 & 6.11 & 0.077 & 0.042 & 0.079 &  0.059 & 0.029 & 0.015 & 0.047\\
7090 & 5.24 & 0.173 & 0.426 & 0.062 &  0.120 & 0.055 & 0.146 & 0.039\\
7095 & 8.74 & 0.122 & 0.148 & 0.134 &  0.040 & 0.052 & 0.054 & 0.085\\
7151 & 1.69 & 0.091 & 0.113 & 0.034 &  0.032 & 0.038 & 0.041 & 0.021\\	
7256 & 10.96 & 0.176 & 0.288 & 0.180 &  0.075 & 0.081 &  0.093 & 0.108\\	
7321 & 3.06 & 0.070 & 0.6900 & 0.061 &  0.011 & 0.027 & 0.222 & 0.034\\
7323 & 3.25 & 0.131 & 0.077 & 0.138 &  0.010 & 0.036 & 0.022 & 0.075\\
7353 & 9.70 & 0.225 & 0.273 & 0.065 &  0.043 & 0.055 & 0.097 & 0.039\\
7524 & 3.58 & 0.102 & 0.087 & 0.083 &  0.002 & 0.033 &  0.028 & 0.051\\
7603 & 2.57 & 0.196 & 0.101 & 0.046 &  0.064 & 0.078 & 0.031 & 0.028\\	
7766 & 13.34 & 0.141 & 0.039 & 0.180 &  0.068 & 0.061 & 0.012 & 0.112\\
7989 & 11.17 & 0.151 & 0.356 & 0.115 &  0.001 & 0.043 & 0.107 & 0.067\\
8863 & 11.53 & 0.107 & 0.195 & 0.081 &  0.009 & 0.042 & 0.058 & 0.049\\
9133 & 42.66 & 0.138 & 0.071 & 0.087 &  0.028 & 0.045 &  0.022 & 0.053\\
9211 & 4.81 & 0.087 & 0.080 & 0.043 &  0.030 & 0.031 & 0.024 & 0.024\\
9649 & 3.59 & 0.151 & 0.081 & 0.116 &  0.002 & 0.057 & 0.024 & 0.074\\	
9858 & 15.16 & 0.380 & 0.250 & 0.183 &  0.089 & 0.132 &  0.077 & 0.111\\
\end{tabular}
$$
\end{table*}
\begin{table*}
$$
\begin{tabular}{p{1.5cm}p{1.5cm}p{1.7cm}p{1.7cm}p{1.7cm}p{1.5cm}p{1.7cm}p{1.7cm}p{1.7cm}}
10359 & 11.49 & 0.137 & 0.226 & 0.181 &  0.013 & 0.067 & 0.079 & 0.119\\
10470 & 13.52 & 0.090 & 0.246 & 0.113 &  0.094 & 0.032 & 0.071 & 0.068\\
11670 & 4.68 & 0.289 & 0.136 & 0.169 &  0.026 & 0.060 & 0.040 & 0.094\\
11707 & 6.98 & 0.094 & 0.178 & 0.051 &  0.069 & 0.022 & 0.045 & 0.031\\
11852 & 31.10 & 0.128 & 0.386 & 0.047 &  0.023 & 0.041 & 0.104 & 0.031\\
11861 & 9.59 & 0.034 & 0.097 & 0.047 &  0.004 & 0.011 & 0.031 & 0.030\\
11891 & 4.44 & 0.203 & 0.054 & 0.096 &  0.021 & 0.072 & 0.017 & 0.059\\
12082 & 4.50 & 0.079 & 0.239 & 0.101 &  0.076 & 0.033 & 0.075 & 0.061\\
12632 & 4.25 & 0.068 & 0.050 & 0.068 &  0.001 & 0.024 & 0.017 & 0.043\\
12732 & 7.28 & 0.106 & 0.092 & 0.120 &  0.036 & 0.043 & 0.032 & 0.074\\
12754 & 3.02 & 0.106 & 0.042 & 0.060 &  0.014 & 0.046 & 0.013 & 0.036\\
\hline
mean & ... & 0.136 & 0.191 & 0.096 & 0.049 & 0.049 & 0.062 & 0.059\\
\hline
\end{tabular}
$$	
\footnotesize {
Notes: (1) galaxy name from the UGC catalogue; (2) Gaussian scale length as measured in this paper; (3) to (5) the mean values of the parameters for morphological lopsidedness averaged between 1 and 2 Gaussian scale lengths; (6) the perturbation parameter in the lopsided potential obtained from the kinematic data in Paper~I; (7) to (9) the mean values of the perturbation parameters obtained from the morphological data in the present paper.}
\end{table*}
\end{appendix}
\end{document}